\documentclass[a4paper,twoside]{article}
\newcommand{\affil}[1]{$^{\rm #1}$}
\usepackage[authoryear]{natbib}
\bibpunct{(}{)}{;}{a}{}{,}
\usepackage{graphicx}
\date{} 
\title{\large\bf\flushleft Malin 1: A Deeper Look}
\author{\parbox{\textwidth}{\flushleft
\vspace{-0.5cm}
{\it L. Moore\affil{A,C} and Q. A. Parker\affil{A,B}}\\
\vspace{0.4cm}
{\small \affil{A}\,Department of Physics, Macquarie University, NSW 2109}\\
{\small \affil{B}\,Anglo-Australian Observatory, PO Box 296, Epping, NSW 2121}\\
{\small \affil{C}\,Email: lmoore@ics.mq.edu.au}}}
\begin{document}
\twocolumn[
\begin{changemargin}{.8cm}{.5cm}
\begin{minipage}{.9\textwidth}
\vspace{-1cm}
\maketitle
%
%
\small{\bf Abstract: Our \textit{R}-band data show that the optical
light from Malin 1 corresponds well with the $>$2-arcmin extent of
the galaxy's HI content and continues well beyond previously
published \textit{V}-band optical light radial profiles. Analysis of
our image yields improved understanding of the galaxy's properties.
We measure ellipticity of 0.20 $\pm $ 0.03, implying inclination of
38 $\pm$ 3$^o$, and we trace the radial profile to 77 arcsec. A
single dusty spiral arm is also weakly discernable, and is
consistent with the rotation direction of the HI and spiral
structure of the inner disk. Possible scenarios for the origin of
the spiral structure are discussed.}


\medskip
\medskip
\end{minipage}
\end{changemargin}
]
\small

\section{Introduction}
\label{intro}
The knowledge of galaxy properties has always been constrained by
limitations of the detector and background noise. By using star
counts from deep CCD imaging, \citet{BH} traced stars in the outer
stellar disk of NGC 300 to 10 scale lengths, doubling the previously
known radial extent of the optical disk. There is still much to by
learned be re-examining objects with improved methods. Here,
traditional astrophotography combined with digital scanning and
stacking enables the outer reaches of the enigmatic galaxy, Malin 1,
to be studied to unprecedented depth. Previous image stacks have
combined up to 13 films \citep{Kat}. Our image is derived from a set
of 63 UK Schmidt Telescope (UKST) films providing a 2.25-mag gain
over the depth of a single film \citep{BH93}. The image is centred
near M87 (the northern concentration of Virgo Cluster galaxies) and
covers an unvignetted field of 25 square degrees. The background
galaxy, Malin 1, also lies within this field.

Malin 1 is a highly unusual galaxy that remains the most extreme in
the class of low surface brightness (LSB) giants \citep{Pick}. A
recent study by \cite{Barth} suggests the galaxy has an inner normal
barred spiral disk (SB0/a) surrounded by a large diffuse LSB
envelope. Nevertheless our data show this galaxy's outer disk
extends to around 120 kpc making it the largest diffuse outer region
of any known disk galaxy. It has a large M$_{HI}$/L$_B$ ratio
\citep{Imp} and appears to be an isolated galaxy that has evolved
slowly. To date, analyses of the properties of Malin 1 have been
hampered by the fact that scaling of the HI rotation curve, and
derivation of dependent quantities, rely on knowing the inclination
of the galaxy \citep{Pick}. The best measurements from CCD images
have large uncertainty in the inclination (e.g. \citet{Pick} quote
inclination of 45 $\pm$ 15$^o$). Improved measurements from our
deeper image allow us to refine this measurement and reduce the
uncertainty considerably.

A summary of previously published data on Malin 1 is included in
Section \ref{comp}. Our data and analysis are described in Section
\ref{image}. In Section \ref{discuss} we consider the significance
of our new findings.

\section{Existing Data}
\label{comp} The discovery of Malin 1 by \citet{Both87} was made by
studying a photographically amplified image. These authors followed
up with deep \textit{V}-band CCD imaging to derive basic optical
parameters for the galaxy, and HI observations with the Arecibo
radio telescope to measure redshift. Malin 1 is acknowledged as
being both massive and dark matter dominated \citep{Pick}. The
published radial profile of \cite{Both87} extends to $\sim$65 arcsec
in \emph{V} band. Their fit to the disk yields an extrapolated
central surface brightness of $V(0)=25.7 \pm 0.1$ mag arcsec$^{-2}$
and disk scale length of 45 $\pm$ 5 arcsec. Table \ref{data}
contains summary results from \citet{Both87} and \citet{Pick}.
Distance-dependent quantities are based on the \citet{Both87} HI
redshift, and assume H$_0$ = 75 km s$^{-1}$ Mpc$^{-1}$ and a derived
distance of 330 Mpc $h^{-1}_{75}$. Values in the table are those of
\cite{Pick} unless otherwise indicated.

\begin{table}[h]
\begin{center}
\caption{Published Data on Malin 1}\label{data}

\begin{tabular}{lcc}
\hline Property & Details$^a$  \\
\hline

RA (1950)& 12$^\textrm{h}$34$^\textrm{m}$28$^\textrm{s}$.1\\
Dec (1950)& +14$^\textrm{o}$36'18".1\\
Redshift & 0.083 $^b$\\
HI velocity & 24 755 $\pm$ 10 km s$^{-1}$\\
Scale length& 45 $\pm$ 5" $^b$\\
Quoted scale size& 82 kpc $h^{-1}_{75}$ \\
Optical radius& 65 arcsec $^b$ \\
Inclination& $45 \pm 15 ^\textrm{o}$  \\
HI diameter& up to 3.3 arcmin $^c$ \\

\hline
\end{tabular}
\medskip\\
$^a$\citet{Pick} unless otherwise stated\\
$^b$\citet{Both87}\\
$^c$\citet{Imp}\\
\end{center}
\end{table}

Measurements of HI mass range between the Impey \& Bothun (1989)
value of 6.7 x 10$^{10}$ M$_\odot$ (for H$_0$=100 km s$^{-1}$
Mpc$^{-1}$) to that of \citet{Rad}, 1.96 x 10$^{11}$ M$_\odot$ (for
H$_0$=75 km s$^{-1}$ Mpc$^{-1}$). Certainly the HI mass is
significant and the M$_{HI}$/L$_B$ ratio is also large at $\sim$ 3
\citep{Imp}. Matthews, van Driel, \& Monner-Ragaigne (2001)
highlight that additional surface photometry and HI measurements are
needed to improve our measurements and understanding of the global
properties of giant LSB galaxies.

\section{Malin 1 Image Data\\ and Analysis}
\label{image} 
\begin{figure}[h]
\begin{center}
\includegraphics[scale=0.4, angle=0]{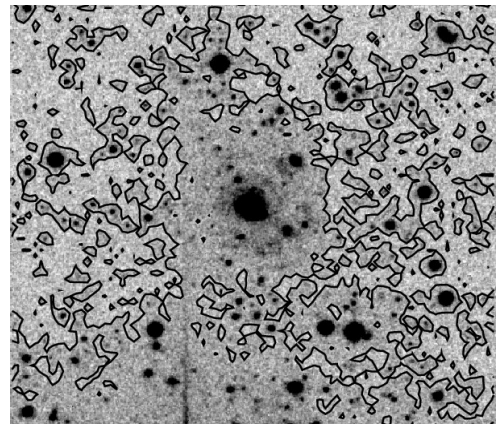}
\caption{Malin 1 with contours at 6380 data counts to delineate the
extent of the optical disk. This image measures 5'36" on a side. The
vertical feature (lower left) is a diffraction spike from a star off
the bottom edge of the image. North is at the top, east is to the
left.}\label{contour6380}
\end{center}
\end{figure}

\begin{figure}[h]
\begin{center}
\includegraphics[scale=0.8, angle=0]{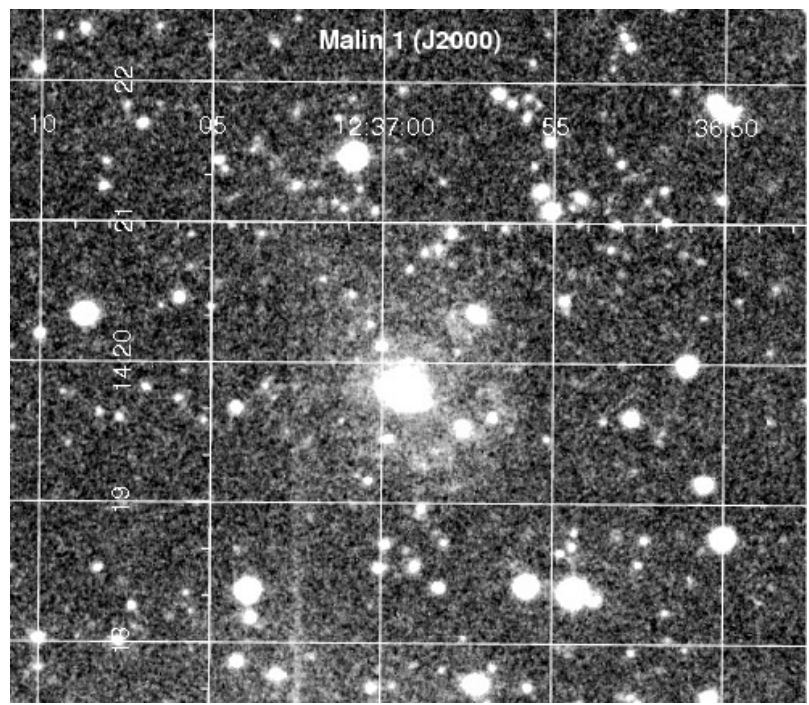}
\caption{Image of Malin 1 from the 63-film stack. Coordinate grid is
J2000.}\label{grid}
\end{center}
\end{figure}

\begin{figure}[h]
\begin{center}
\includegraphics[scale=1.0, angle=0]{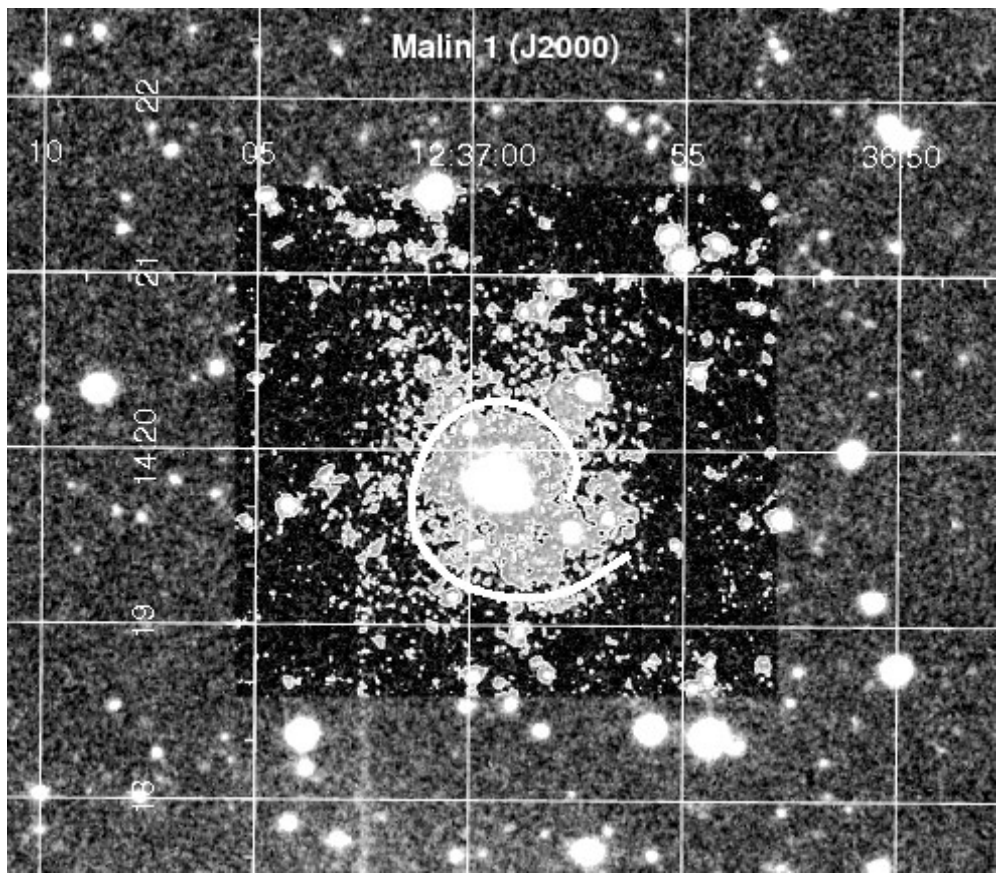}
\caption{Malin 1 with dark and light contours overlaid to enhance
intensity variations in the image. The drawn line traces the
position of the dusty spiral arm detected in the image.}\label{line}
\end{center}
\end{figure}

The image of Malin 1 (base image of Figures
\ref{contour6380}, \ref{grid} and \ref{line}) is derived from 63
UKST films that have been digitally scanned by SuperCOSMOS and
co-added to produce a unique ultra-deep \textit{R}-band image
covering 36 square degrees. The 63 A-grade Tech--Pan films were
taken with an OG590 filter between January and June in 1999, 2000
and 2001. Exposure times were mostly 60.0 minutes, though 13 were a
little shorter, probably due to interference by cloud. The field
centre is at 12$^\textrm{h}$ 27$^\textrm{m}$ 00$^\textrm{s}$
+13$^\textrm{o}$ 30" 00' (B1950). This places Malin 1 in the upper
left portion of the UKST field, though not so close to the edge as
to be affected by vignetting. The range in data counts is optimised
for the faint end in the scanning process, so bright stars and
galaxy cores are affected by saturation.

The stacking of films increases signal to noise, having the effect
of pushing down the background to reveal fainter parts of the galaxy
than are visible in a single film. The magnitude gain from stacking
$N$ films is given by \cite{BH93} as 2.5 log $\surd {N}$ magnitudes.
Hence, the deep-stack image goes 2.25 magnitudes deeper than a
single Tech--Pan film, which is already one magnitude deeper than
traditional IIIa emulsions used on glass plates \citep{PM}.

Calibration of the image was performed by B. J. Jones (2006, private
communication) by selecting 20 stars in the magnitude range 18.2
$\leq$ Cousins \textit{R} $\leq$ 20.1 from the Sloan Digital Sky
Survey DR4 online catalogue and using aperture photometry on the
same stars in the image. A best fit to the magnitudes yields a
calibration constant that can then be applied to the conversion
between data counts and magnitude anywhere on the image. The
calibration constant thus found is 31.54.

\begin{figure}
\begin{center}
\includegraphics[scale=0.8, angle=0]{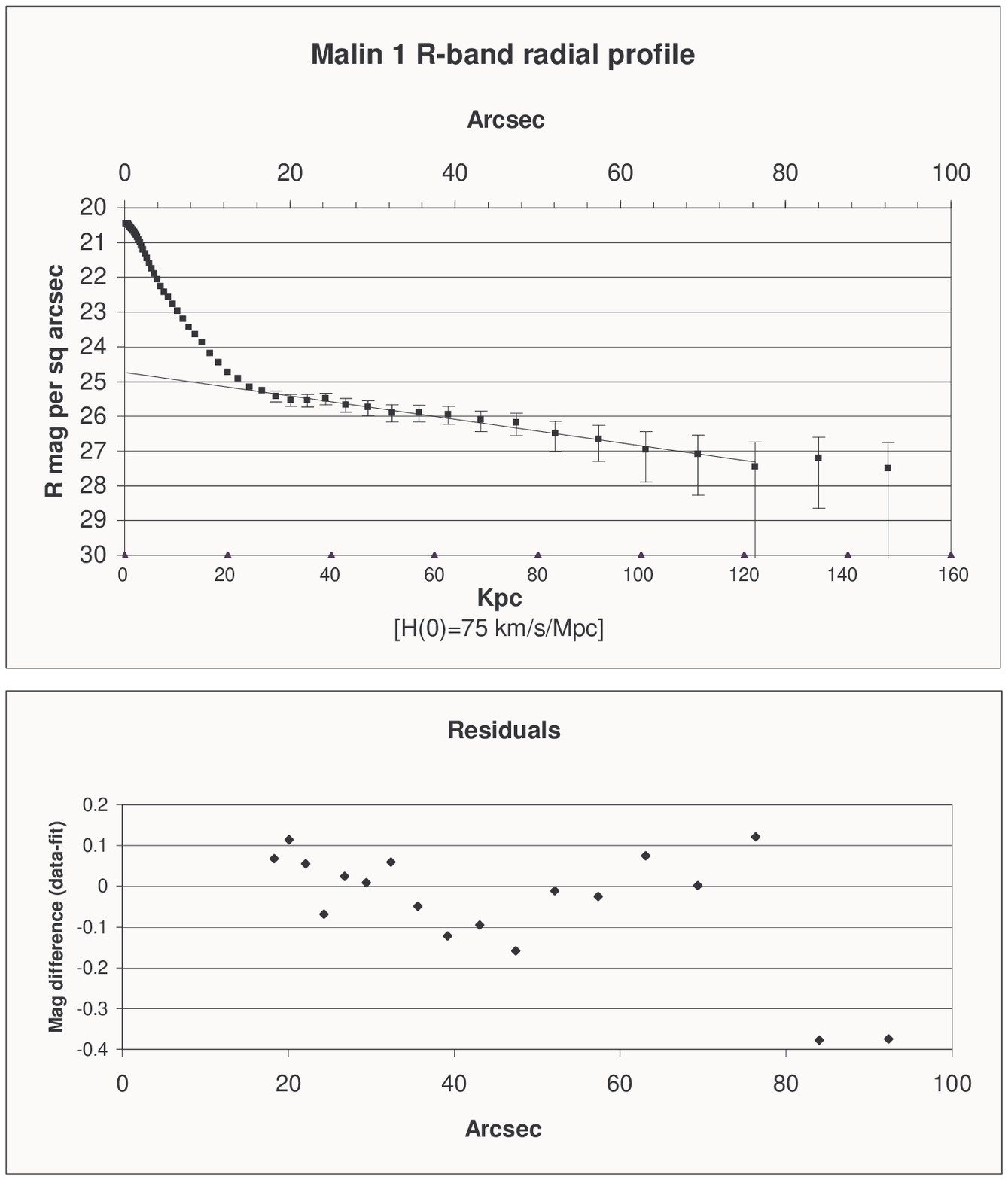}
\caption{\emph{Top}: The \textit{R}-band radial profile shows that
the optical disk (solid squares) is visible to a semi-major axis of
$\sim$ 80 arcsec before reaching the background level. The upper
scale shows arcsec. The lower scale shows kpc based on the redshift
give in Table 1 and assuming H$_0$ = 75 km s$^{-1}$ Mpc$^{-1}$. A
fit to the exponential part of the profile is shown as a solid line.
The last two data points on the right are not included in the fit as
these are clearly beyond the point where the galaxy light can be
distinguished from sky and adjacent objects. The fit is extrapolated
to the left to show central surface brightness of 24.74 \emph{R} mag
arcsec$^{-2}$. See text for explanation of the calculation of the
error bars on the exponential part of the profile.
\newline
\emph{Bottom}: Residuals from the fit (data minus fit), also showing
the variance of the last two data points from the fit.}
\label{profile}
\end{center}
\end{figure}

Our data show that the \textit{R}-band optical light from Malin 1
continues even further than the HI mapped by \citet{Pick} and
reproduced by Braine, Herpin, \& Radford (2000). Contours at 6380
data counts on the image are displayed in Figure \ref{contour6380}
and trace the irregular outer extent of the galaxy. We also find the
radial profile (Figure \ref{profile}) extends to almost 80 arcsec,
well beyond the \textit{V}-band profiles of \citet{Both87} and
\citet{Pick}. The \cite{Both87} profile only extends to $\sim$65
arcsec, and the \cite{Pick} profile is plotted to $\sim$80 kpc,
equating to approximately 50 arcsec (where they have used H$_0$ = 75
km s$^{-1}$ Mpc$^{-1}$).

Ellipse parameters of the outer isophotes of Malin 1 were determined
by matching ellipses by eye to contours overlaid on the image at
intervals of 20 data counts in the range 6360 to 6460 inclusive.
These were displayed in \textsc{ds9} and ellipse dimensions and
position angles read off. The mean values thus derived are
ellipticity $e=0.20 \pm 0.03$ ($e = 1-b/a$ where $a$ and $b$ are
major and minor axes respectively), and position angle $\phi = 43
\pm 3^o$. The former equates to an inclination of 38 $\pm$ 3$^o$.
Combinations of these parameters were used as fixed values for
surface photometry in the \textsc{iraf} task \textsc{ellipse} with
resulting profiles yielding only small variations in extrapolated
central surface brightness and scale length.

The radial profile from the surface photometry using ellipticity of
0.20 and position angle of 40 degrees is shown in Figure
\ref{profile}. Note that the outer parts of the radial profile are
highly sensitive to the background value used. The sky background
level was measured in five carefully selected separate regions from
2.5 to 3 arcmin away from the galaxy's optical centre. These
elliptical regions, arranged on three sides of the galaxy and each
covering $\sim$2100 pixels, were evaluated in \textsc{gaia} to
obtain separate mean and standard deviation for each. The five
results were then averaged to arrive at a final mean of 6364 and
standard deviation of 38 data counts.


The radial profile (Figure \ref{profile}) may be traced to 77 arcsec
on the semi-major axis of the outermost fitted ellipse before
levelling out to the sky background. This is equivalent to 2.3 $\pm$
0.1 scale lengths or 124 kpc $h^{-1}_{75}$. This does not include an
extension to the north-east, which does appear from our image to be
contiguous with Malin 1, but could be a tidal tail, an interacting
or completely separate object.

The largest source of error in the outer part of the profile is due
to uncertainty in the sky background measurement. Ideally the errors
on the data points in the profile should be derived by summing in
quadrature the errors in signal and sky. Unfortunately, this fails
when attempting to convert a resulting negative value (where signal
minus error is less than sky) to magnitude. Instead, errors here are
derived from recalculations of magnitudes using: sky$_{high/low}$ =
sky$_{mean} \pm (0.5\sigma_{sky}$).

The fit to the exponential part of the radial profile yields a value
of 24.74 (+0.07/-0.2) \textit{R} mag arcsec$^{-2}$ for extrapolated
central surface brightness, with errors taken from extrapolated fits
to error bars. Scale length of the fit is measured at 33 arcsec (53
kpc $h^{-1}_{75}$), which is smaller than the values obtained by
either \citet{Both87} or \citet{Pick}. We also find a variation in
position angle of fitted ellipses that may support the warping
mentioned by \citet{Pick}.

Inspection of the image at high contrast levels (Figure \ref{grid})
and enhancement using overlaid contours reveal for the first time
the likely presence of a single dusty spiral arm. \cite{Barth}
comments that spiral structure is often seen in outer disks of
early-type barred galaxies and he detects a hint of spiral structure
in his HST \textit{I}-band image of Malin 1. Clearly, it is
difficult to discern such structure at this magnitude where contrast
with the background is so low. For the reader's benefit, contours
and the position of the spiral arm are drawn onto Figure \ref{line}.
Inspection of a high-resolution jpeg version of David Malin's
\textit{B}-band image
(http://www.aao.gov.au/images/deep$\_$html/malin-1$\_$d \\.html)
shows tonal variations consistent with our measurements.

\section{Discussion}
\label{discuss}

The apparent presence of a single dusty spiral arm has some profound
implications. Asymmetric spiral formation may be a natural
consequence of the lack of a gaseous damping process \citep{Phook},
which stands to reason when there is low gas surface density. The
shape of the spiral arm is consistent with the rotation direction of
the HI mapped by \cite{Pick} and the detailed view of the inner disk
studied by \cite{Barth}. Modelling of the formation of spiral arm
structure is beyond the scope of this paper, but our detection poses
the question: Could a single spiral arm evolve naturally in a giant
quiescent disk galaxy?

In a previous study of a one-armed spiral galaxy \cite{Phook} imply
that this type of structure is more likely the result of an
interaction with an infalling gas cloud. Modelling by
Pe{\~n}arrubia, McConnachie, \& Babul (2006) shows that a diffuse
outer disk could result from tidal shredding of a dwarf galaxy. If
this occurred and was relatively recent, it could account for the
observed asymmetry. However, theories on the nature of LSB galaxies
suggest that star formation is inhibited by low gas surface density
\citep{Kenn} and a quiescent history without interactions. Therefore
it seems that if the latter idea is correct, the single-arm
structure should not be due to an interaction.

On the other hand, from our image we also discern a faint optical
lobe to the north-east (near top, left of centre in Figure
\ref{contour6380}). Whilst it is impossible to say, without further
detailed observation, whether or not this is associated with Malin
1, we conjecture that this may be part of a system undergoing
`quiescent merger'. This may explain both the single-arm structure,
as material is gently accreting onto the galaxy, and the observed
low surface brightness, maintained through lack of shocks that might
otherwise trigger star formation. We note that the optical light we
detect for the galaxy itself corresponds directly with the HI map of
\cite{Pick}, extending across more than 2 arcmin. Though the HI
mapped by \cite{Pick} does not appear to extend into the north-east
lobe region, their data show some extension in that direction.

If the galaxy is undergoing a quiescent merger, it is likely that
this process is extremely rare and may depend on the unique
properties of Malin 1, i.e. its M$_{HI}$/L$_B$ of $\sim$ 3
\citep{Imp}, its high dark matter content \citep{Pick} and isolation
from neighbouring galaxies.

\section*{Acknowledgments} 
We wish to thank Bryn Jones for his calibration of the image stack
and his helpful private communications. We also thank David Malin
for the opportunity to study his \textit{B}-band image of Malin 1.

\onecolumn


\twocolumn

\end{document}